\documentclass[aps,pra,floatfix,amsmath,amssymb,showpacs,showkeys,twocolumn,10pt]{revtex4-1}
\usepackage[caption=false]{subfig}
\usepackage{graphicx,bm,color}
\usepackage{amsfonts}
\usepackage{epstopdf}
\begin{document}

\frenchspacing

\title{Higher order quantum vortex}

\begin{abstract}
In this article we present a quantum theoretical framework of photons entangled in higher order modes like Ince-Gauss modes. Ince-Gauss modes are the natural solution of the quantum harmonic oscillator in elliptical coordinates. These modes are defined by an additional quantum number, the ellipticity parameter which is expected to play a major role in continuous variable quantum key distribution protocols. Although such quantum states have been realized in experiments but a proper theoretical framework has been missing. We aim to address this shortcoming with our article.
\end{abstract}

\author{Anindya Banerji$^{1,2}$}\email[Corresponding author: ]{abanerji09@gmail.com}
\author{Ravindra P. Singh$^{3}$}\email{rpsingh@prl.res.in}
\author{Abir Bandyopadhyay$^{2}$}\email{abir@hetc.ac.in}
\affiliation{$^{1}$Department of Physics, Jadavpur University, Kolkata 700032, India}
\affiliation{$^{2}$Hooghly Engineering and Technology College, Hooghly 712103, India}
\affiliation{$^{3}$Physical Research Laboratory, Ahmedabad 380009, India}

\date{\today}

\pacs{42.65.Lm, 03.65.Ud, 03.67.Mn, 03.67.Bg, 89.70.Cf}
\keywords{Ince-Gaussian modes, Wigner function, Entanglement, Elliptical coordinates}

\maketitle

\section{Introduction}
Quantum states characterised by an azimuthal index or topological charge are called quantum vortex states. Such states of the radiation field was introduced in \cite{GSA97}. These states exhibit singularities in the phase space. The number of singularities are dependent on the azimuthal index of the vortex, also known as the order of the vortex. Quantum vortex states are usually described by Laguerre-Gaussian modes in the quadrature space. They have a central region of zero intensity bounded by a region of uniform intensity. The radius of the boundary is also dependent on the order of the vortex. Higher the order, greater is the radius. These Laguerre-Gauss vortex states have been subjected to detailed study over the last few years \cite{jbvortex,abir,GSANJP,JPhysA,OptComm1,OptComm2}. Also, quantum vortex states described by Hermite-Gaussian and Bessel-Gaussian distribution \cite{BGV} have been investigated. We had introduced the perfect quantum optical vortex state \cite{PerfVort} in which the radius of the central dark core remains unchanged even with increasing order of the vortex. The continued interest in the study of these states stems from the fact that these are highly entangled states carrying orbital angular momentum (OAM). When expressed in the OAM basis, these states exist in an infinite Hilbert space.\\
The state that we consider in this article exhibits a complex vortex nature. We call it the higher order quantum vortex state. In the spatial quadratures, it resembles a helical Ince-Gaussian (HIG) mode decomposition. Ince-Gaussian mode solutions of the stable oscillators were introduced some time back \cite{Bandres2004, Bandres2004_2, Bandres2008}. These mode functions consist of the Ince polynomials which are named after E. G. Ince who introduced them in 1923 \cite{Ince}. These polynomials are solutions of the general Hill equation and have been studied in detail in \cite{Arscott, Arscott2}. Quite recently, photons entangled in helical Ince-Gauss modes were also studied experimentally \cite{IG_Zeilinger}.\\
The HIG modes that describe the higher order quantum vortex we study here are defined using two discrete quantum numbers; the order $p$ and degree $m$. Additionally, there is a third continuous parameter called the ellipticity, $\epsilon$. Each value of this continuous parameter gives rise to a family of orthogonal Ince-Gaussian modes. Each Ince-Gaussian mode can be further classified into two groups; the even and odd Ince-Gaussian modes. The importance of these exotic modes lies in the fact that they form the third complete family of solutions of the quantum harmonic oscillator in elliptical coordinates. Further, they constitute a continuous transition basis between the more common Hermite-Gaussian modes and (even and odd) Laguerre-Gaussian modes. The even/odd Ince-Gaussian modes reduce to Hermite-Gaussian modes when $\epsilon \rightarrow\infty$ and even/odd Laguerre-Gaussian modes when $\epsilon \rightarrow 0$. Therefore, the Ince-Gaussian modes represent a more general form of solutions of the quantum harmonic oscillator.\\
This article is organised as follows. In section \ref{sec:Theoretical generation}, we outline the method for theoretical generation of the higher order quantum vortex states exhibiting an Ince-Gaussian distribution in the quadrature space. We calculate the Wigner distribution function associated with the higher order vortex states in section \ref{sec:Entanglement and nonclassical properties}. We also study the entanglement between the two modes using von Neumann entropy in the same section. We summarise the main results of the article in section \ref{sec:Conclusion}.
\begin{figure*}
\centering
\subfloat[p=3,m=3,$\epsilon$=2]{
\includegraphics[scale=0.25]{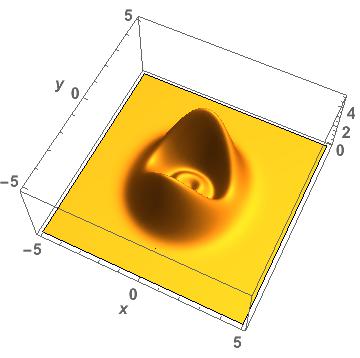}
\label{fig:IG332}}
\qquad
\subfloat[p=5,m=3,$\epsilon$=2]{
\includegraphics[scale=0.25]{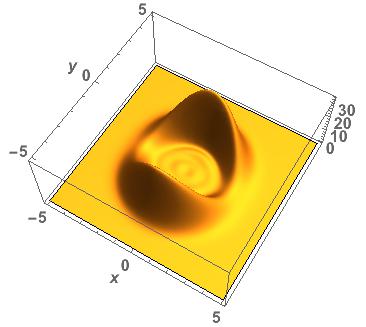}
\label{fig:IG532}}
\qquad
\subfloat[p=7,m=3,$\epsilon$=2]{
\includegraphics[scale=0.25]{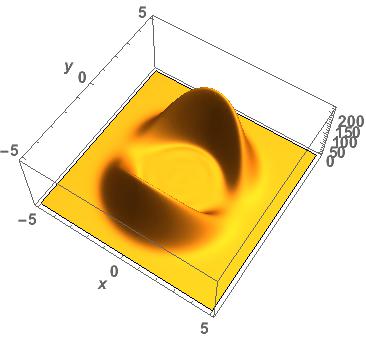}
\label{fig:IG732}}
\qquad\\
\subfloat[p=5,m=1,$\epsilon$=2]{
\includegraphics[scale=0.25]{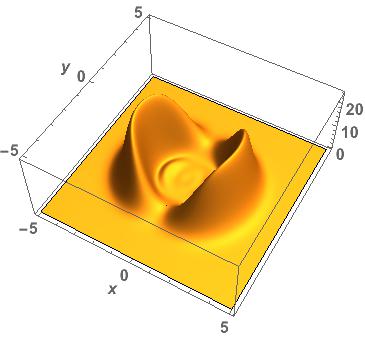}
\label{fig:IG512}}
\qquad
\subfloat[p=5,m=3,$\epsilon$=2]{
\includegraphics[scale=0.25]{IG532.jpg}
\label{fig:IG532a}}
\qquad
\subfloat[p=5,m=5,$\epsilon$=2]{
\includegraphics[scale=0.25]{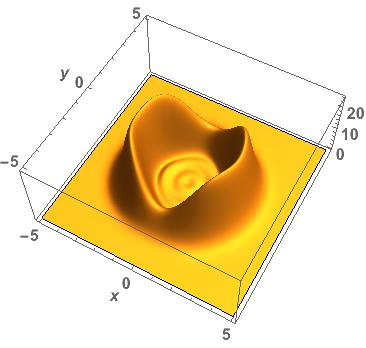}
\label{fig:IG552}}
\qquad\\
\subfloat[p=3,m=3,$\epsilon$=0]{
\includegraphics[scale=0.25]{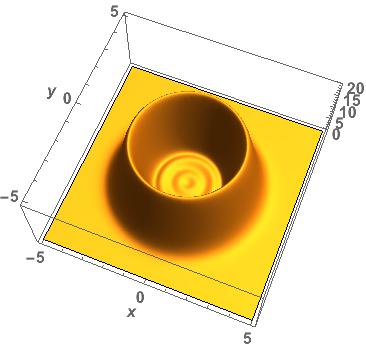}
\label{fig:IG330}}
\qquad
\subfloat[p=3,m=3,$\epsilon$=5]{
\includegraphics[scale=0.25]{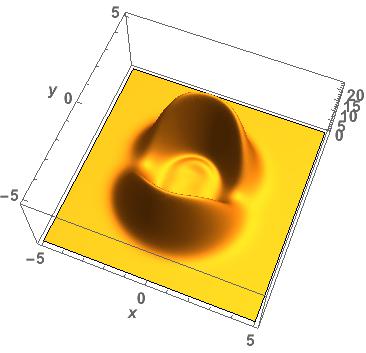}
\label{fig:IG335}}
\qquad
\subfloat[p=3,m=3,$\epsilon=\infty$]{
\includegraphics[scale=0.25]{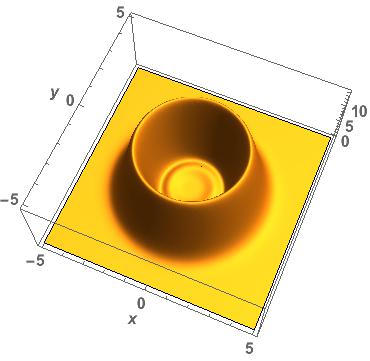}
\label{fig:IG33Infinity}}
\caption{Intensity distribution of Ince-Gaussian vortex states for various orders and degrees. Details are in the text.}
\label{fig:IGQuadratures}
\end{figure*}
\section{Theoretical generation}
\label{sec:Theoretical generation}
The family of Ince-Gauss mode solutions of the paraxial wave equation can be broadly divided into two families: even and odd Ince-Gauss modes. These can be represented in the closed form as
\begin{eqnarray}
\label{InceGauss}
IG_{p,m}^{e}\left(\mathbf{r},\epsilon\right)&=&N_eC_p^m\left(i\xi,\epsilon\right)C_p^m\left(\eta,\epsilon\right)\exp\left(\frac{-r^2}{2}\right)\\
IG_{p,m}^{o}\left(\mathbf{r},\epsilon\right)&=&N_oS_p^m\left(i\xi,\epsilon\right)S_p^m\left(\eta,\epsilon\right)\exp\left(\frac{-r^2}{2}\right)
\end{eqnarray}
where $r^2=x^2+y^2$ and $\epsilon$ is the ellipticity parameter. $C_p^m$ and $S_p^m$ are the even and odd Ince polynomials of order $p$ and degree $m$. Both $p$ and $m$ has the same parity, i.e. $(-1)^{(p-m)}=1$ and takes values $1\leq m\leq p$ for odd polynomials and $0\leq m \leq p$ for even polynomials. The spatial quadrature coordinates are related to the elliptic coordinates $\xi$ and $\eta$ as follows
\begin{equation}
\begin{pmatrix} x\\y \end{pmatrix} = \epsilon\begin{pmatrix} \cosh \xi \cos \eta \\ \sinh \xi \sin \eta \end{pmatrix}
\end{equation}
A helical Ince-Gaussian state can be decomposed in terms of the odd and even Ince-Gaussian modes and is written as
\begin{equation}
\label{HelicalIG}
\vert HIG\rangle_{p,m}=\vert IG\rangle_{p,m}^e \pm i\vert IG\rangle_{p,m}^o
\end{equation}
\noindent where $\vert IG\rangle_{p,m}^{\sigma}$ represents Ince-Gaussian mode of the corresponding polarity $\sigma$. The Ince-Gaussian modes can be expanded in odd/even Laguerre-Gauss basis as follows
\begin{equation}
\label{IGinLG}
\vert IG\rangle_{p,m}^{\sigma}=\sum_{n}A_{n,l}^{\sigma}\vert LG\rangle_{n,l}^{\sigma}
\end{equation}
\noindent where $n$ is the radial index and $l$ is the azimuthal index. The order $p$ of the Ince-Gaussian mode, radial index $n$ and azimuthal index $l$ are related through $p=2n+l$. $A_{n,l}^{\sigma}$ are normalized expansion coefficients for a given parity and can be calculated with the help of the overlap integral. The odd/even Laguerre-Gauss modes have the following form
\begin{eqnarray}
\label{LGEvenOddExpansion}
LG_{n,l}^{e,o}&=&\sqrt{\frac{4n!}{\pi(n+l)!}}\left(\sqrt{2}r\right)^{l}\begin{pmatrix} \cos l\phi \\ \sin l\phi\end{pmatrix} \nonumber \\
&\times & L_n^l\left(2r^2\right)\exp\left(\frac{-r^2}{2}\right)
\end{eqnarray}
\noindent where  $L_n^l\left(2r^2\right)$ is the Laguerre polynomial with radial index $n$ and azimuthal index $l$ and $r^2=x^2+y^2$. Now, the quantum state of photons carrying orbital angular momentum are written in terms of the Laguerre polynomials with an additional azimuthal phase term. These quantum optical vortex states have the following form in the spatial quadratures
\begin{eqnarray}
\label{LGquadrature}
\psi\left(r,\phi\right)_{LG}&=&\sqrt{\frac{4n!}{\pi(n+l)!}}\left(\sqrt{2}r\right)^le^{il\phi}\nonumber \\
&\times & L_n^l\left(2r^2\right)\exp\left(\frac{-r^2}{2}\right)
\end{eqnarray}
where $\left(r,\phi\right)$ are polar coordinates. It is easy to see that Eq. (\ref{LGquadrature}) can be written in as a combination of odd and even Laguerre-Gauss modes using Eq. (\ref{LGEvenOddExpansion}). With this necessary background, we propose a quantum optical vortex state with photons exhibiting a helical Ince-Gaussian mode distribution in the spatial quadratures. It can be written as a superposition of Laguerre-Gaussian modes in the following way
\begin{equation}
\label{IGVortex}
\vert IGV\rangle_{p,m}=\sum_{n}A_{n,l}\vert LG\rangle_{n,l}
\end{equation}
where $A_{n,l}$ are normalized weight factors corresponding to each term in the expansion and $\vert LG\rangle_{n,l}$ describe photons carrying orbital angular momentum $l\hbar$. The indices $p$, $n$ and $l$ are related same as before. Now let us consider a two mode state carrying $N$ photons in total. In the Fock basis, we can write it as
\begin{equation}
\label{Initial}
\vert\psi\rangle=\sum_{j}^{\frac{N-1}{2}} A_j\vert N-j,j\rangle
\end{equation}
It might be noted that any two mode quantum state of $N$ particles can be written in the form of Eq. (\ref{Initial}) in the Fock basis. The only difference being that we consider a truncated sum of only half the number of terms than in a general expression. The reason for this will be cleared in a while.\\
The spatial quadrature distribution associated with Eq. (\ref{Initial}) resembles a Hermite-Gaussian mode distribution. In order to convert it to a Laguerre-Gaussian mode distribution with an azimuthal phase, we need to define a mode conversion operator. It terms of the bosonic creation and annihilation operators, it has the form
\begin{equation}
\label{ModeConverter}
\hat{C}=\frac{1}{2}\left(a^{\dagger}b+ab^{\dagger}\right)
\end{equation}
Under the action of this operator, Eq. (\ref{Initial}) evolves under a unitary transformation to the following
\begin{equation}
\label{LG}
\vert\psi\rangle_{\hat{C}}=\exp\left(i2\phi\hat{C}\right)\vert\psi\rangle
\end{equation}
This means that Eq. (\ref{ModeConverter}) rotates the $\left(x,p_y\right)$ and $\left(y,p_x\right)$ planes by an angle $\phi$. It is common knowledge that a $\pi/4$ converter is employed in experiments to transform Hermite-Gaussian mode to a Laguerre-Gaussian mode with azimuthal phase dependence. Using $\phi=\pi/4$ in Eq. (\ref{LG}), we get the following
\begin{equation}
\label{LGV}
\vert\psi\rangle_{v}=\exp\left(i\frac{\pi}{2}\hat{C}\right)\vert\psi\rangle
\end{equation}
In the Heisenberg picture, this translates to the following
\begin{equation}
\label{IGV}
\vert\psi\rangle_{v}=\sum_j^{\frac{N-1}{2}}A_j\left(\frac{a^{\dagger}}{\sqrt{2}}+\frac{ib^{\dagger}}{\sqrt{2}}\right)^{N-j}\left(\frac{b^{\dagger}}{\sqrt{2}}+\frac{ia^{\dagger}}{\sqrt{2}}\right)^j\vert 0,0\rangle
\end{equation}
Solving for the operators, we arrive at the final closed form
\begin{eqnarray}
\label{FinalClosedForm}
\vert\psi\rangle_{v}&=&\sum_{j=0}^{\frac{N-1}{2}}A_j\sqrt{\frac{j!(N-j)!}{2^N}}\nonumber\\
&\times &\sum_{k=0}^{j}\sum_{l=0}^{N-j}c_{lk}^{Nj}\vert N-(j+l-k),j+l-k\rangle
\end{eqnarray}
where $A_j$ are the normalized expansion coefficients and $c_{lk}^{Nj}$ are defined as follows
\begin{equation}
\label{coefficients}
c_{lk}^{Nj}=\frac{i^{k+l}\sqrt{(N-j-l+k)!(j+l-k)!}}{k!(j-k)!l!(N-j-l)!}
\end{equation}
The Eq. (\ref{FinalClosedForm}) can be simplified further and written in the Laguerre-Gaussian basis as follows
\begin{equation}
\label{LGBasis}
\vert\psi\rangle_{v}=\sum_j^{\frac{N-1}{2}}A_j\vert LG\rangle_{j,N-2j}
\end{equation}
where $\vert LG\rangle_{j,N-2j}$ represents Laguerre-Gaussian mode with radial index $j$ and azimuthal index $N-2j$. Comparing Eq. (\ref{LGBasis}) with Eq. (\ref{IGVortex}), it is easy to see that Eq. (\ref{FinalClosedForm}) describes an Ince-Gaussian vortex state in the Fock basis which we call the higher order quantum vortex. The expansion coefficients $A_j$ can then be derived from the overlap integral as follows
\begin{eqnarray}
\label{Overlap}
A_j&=&\int\int_{-\infty}^{\infty}\left(IG_{p,m}^e+iIG_{p,m}^o\right)\nonumber\\
&\times &\left(LG_{j,N-2j}^e+iLG_{j,N-2j}^o\right)\text{d}S
\end{eqnarray}
As an example, for $p=N=5$, $m=1$ and $\epsilon=2$, the corresponding Ince-Gaussian vortex state is written in terms of Laguerre-Gaussian modes as follows
\begin{equation}
\label{Numerical}
\vert\psi\rangle_v^{5,1,2}=0.2079\vert LG\rangle_{0,5}+0.8279\vert LG\rangle_{1,3}-0.5210\vert LG\rangle_{2,1}
\end{equation}
\begin{figure*}
\centering
\subfloat[$W(x,y)_{p_x=0,p_y=0}$]{
\includegraphics[scale=0.25]{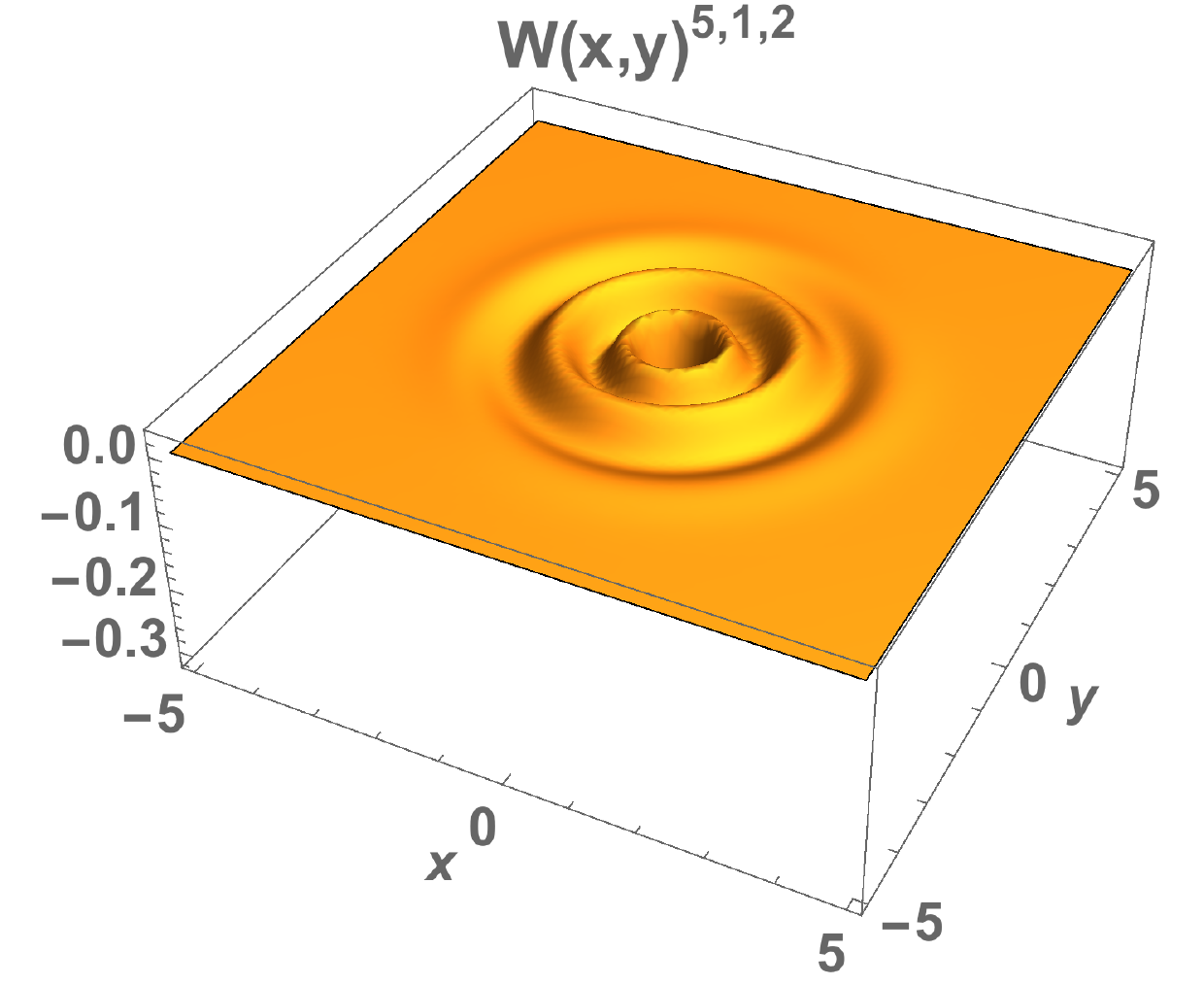}
\label{fig:WXY}}
\qquad
\subfloat[$W(x,p_x)_{y=0,p_y=0}$]{
\includegraphics[scale=0.25]{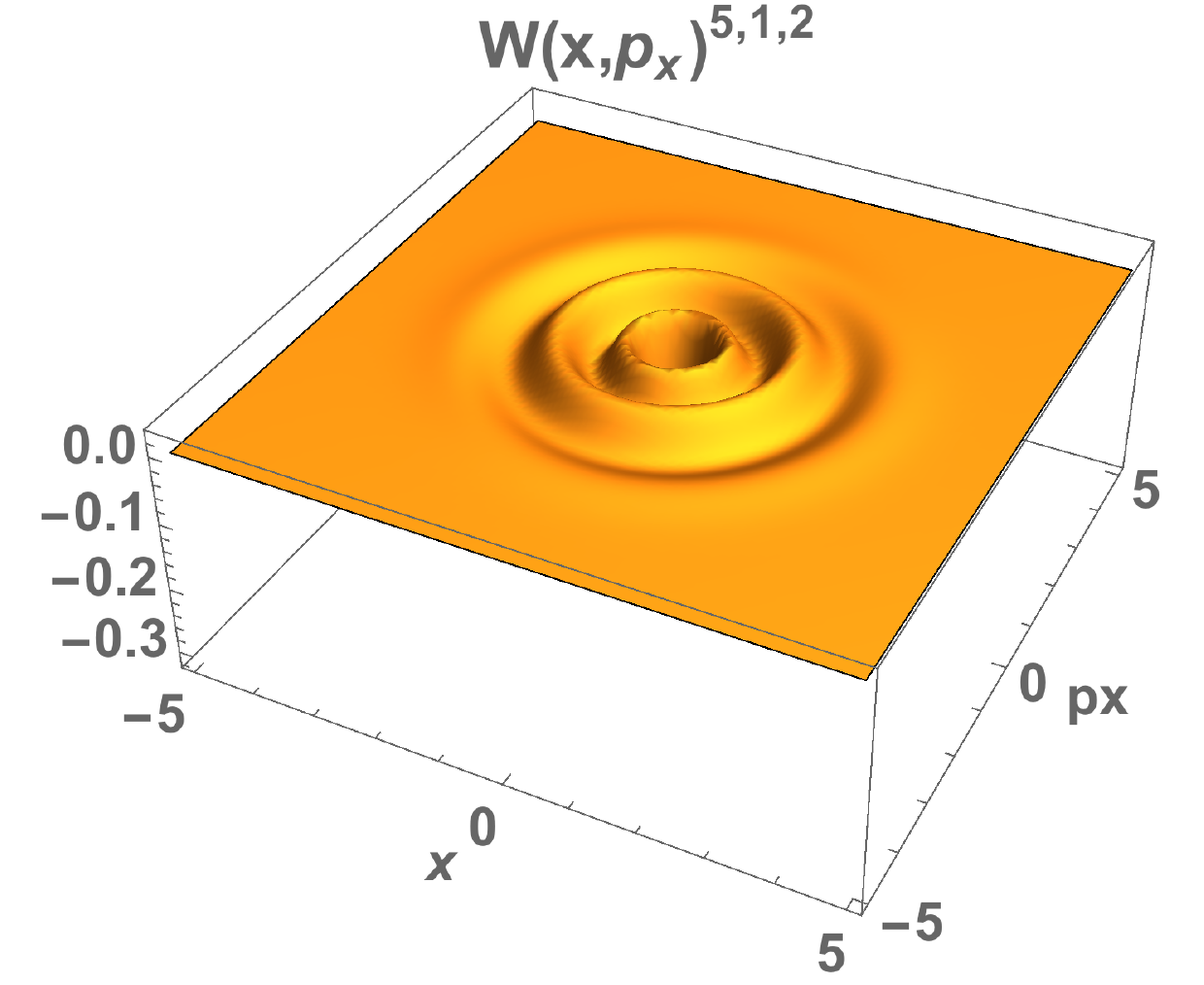}
\label{fig:WXPx}}
\qquad
\subfloat[$W(x,p_y)_{y=0,p_x=0}$]{
\includegraphics[scale=0.25]{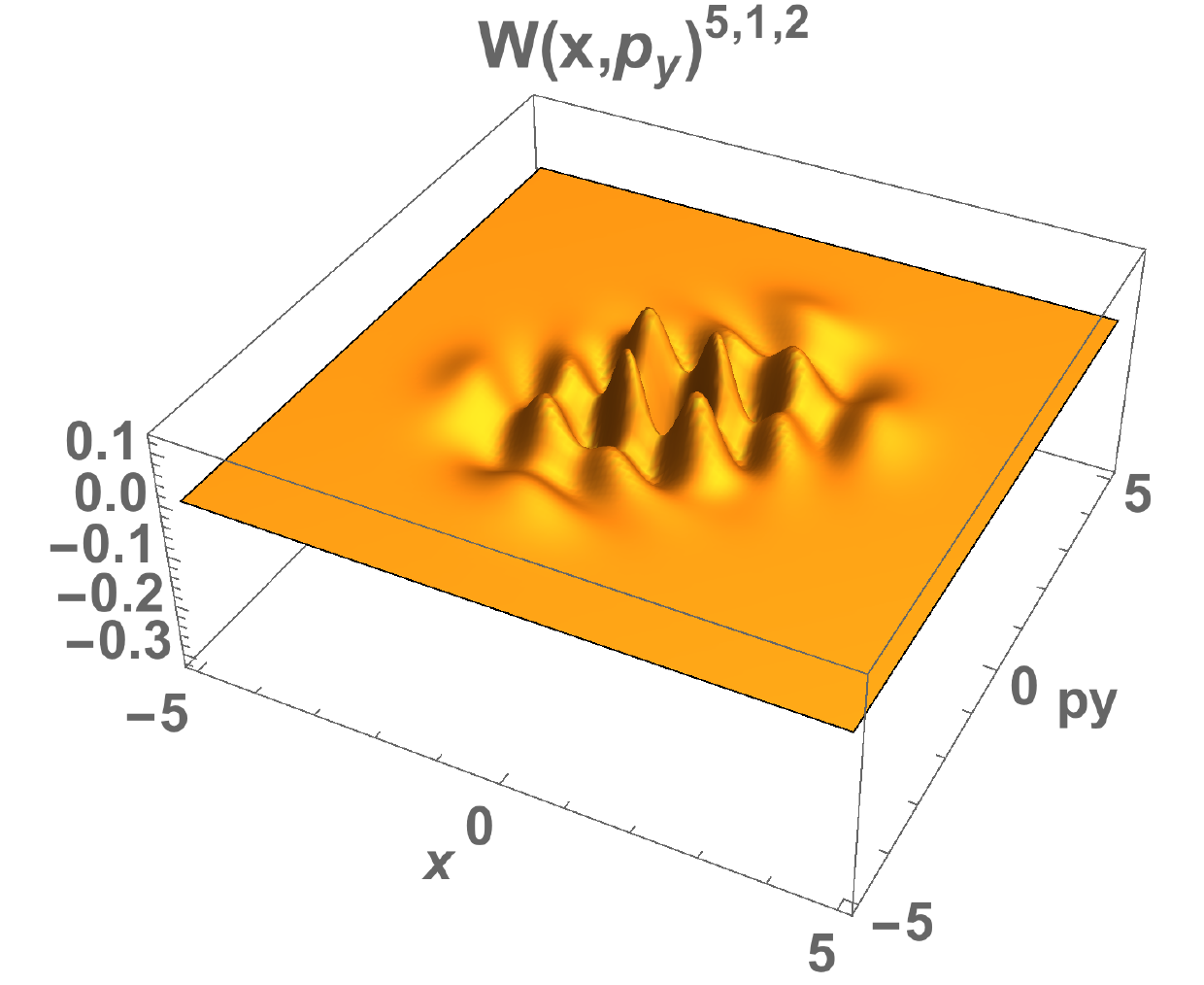}
\label{fig:WXPy}}
\qquad\\
\subfloat[$W(p_x,p_y)_{x=0,y=0}$]{
\includegraphics[scale=0.25]{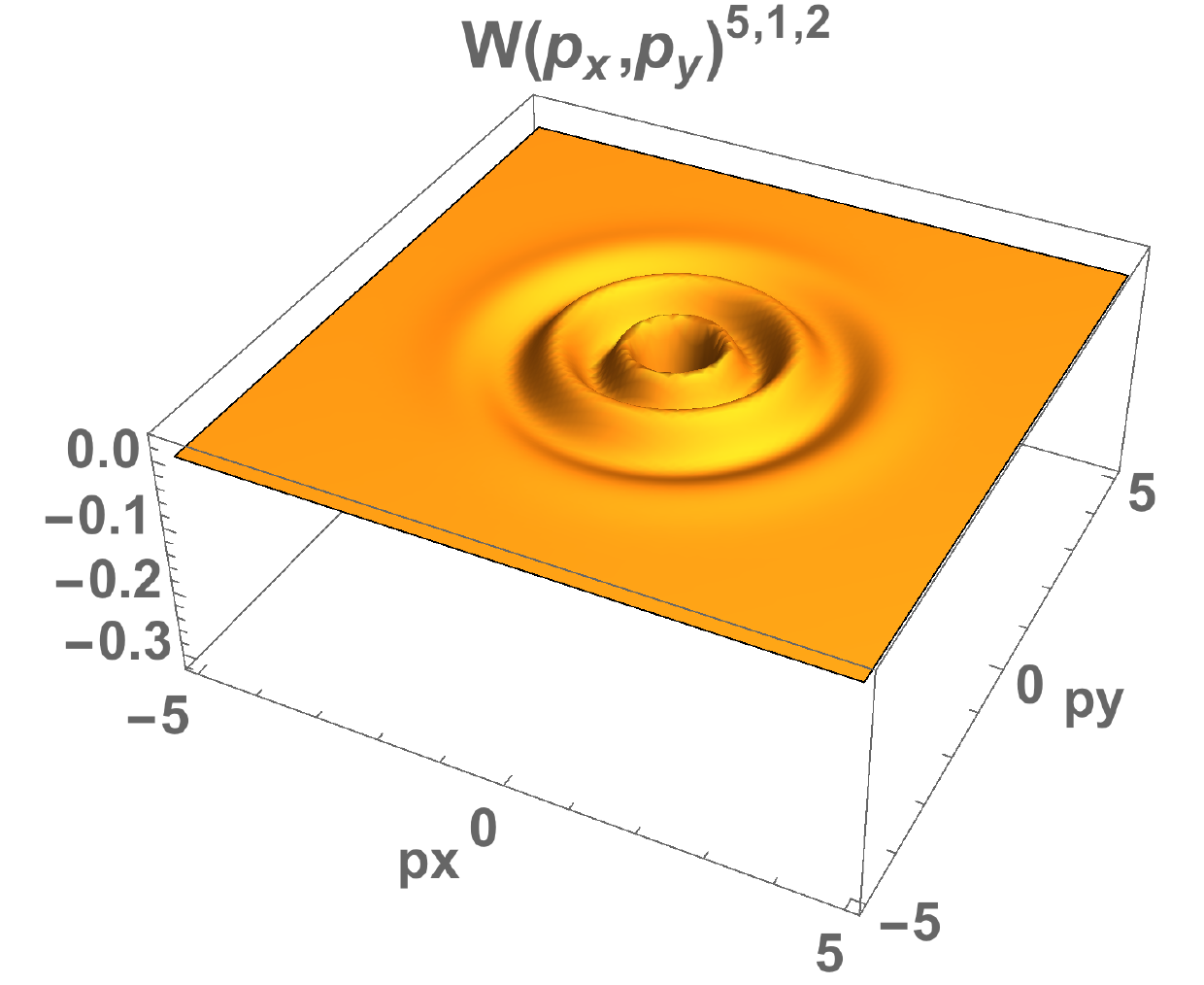}
\label{fig:WPxPy}}
\qquad
\subfloat[$W(y,p_y)_{x=0,p_x=0}$]{
\includegraphics[scale=0.25]{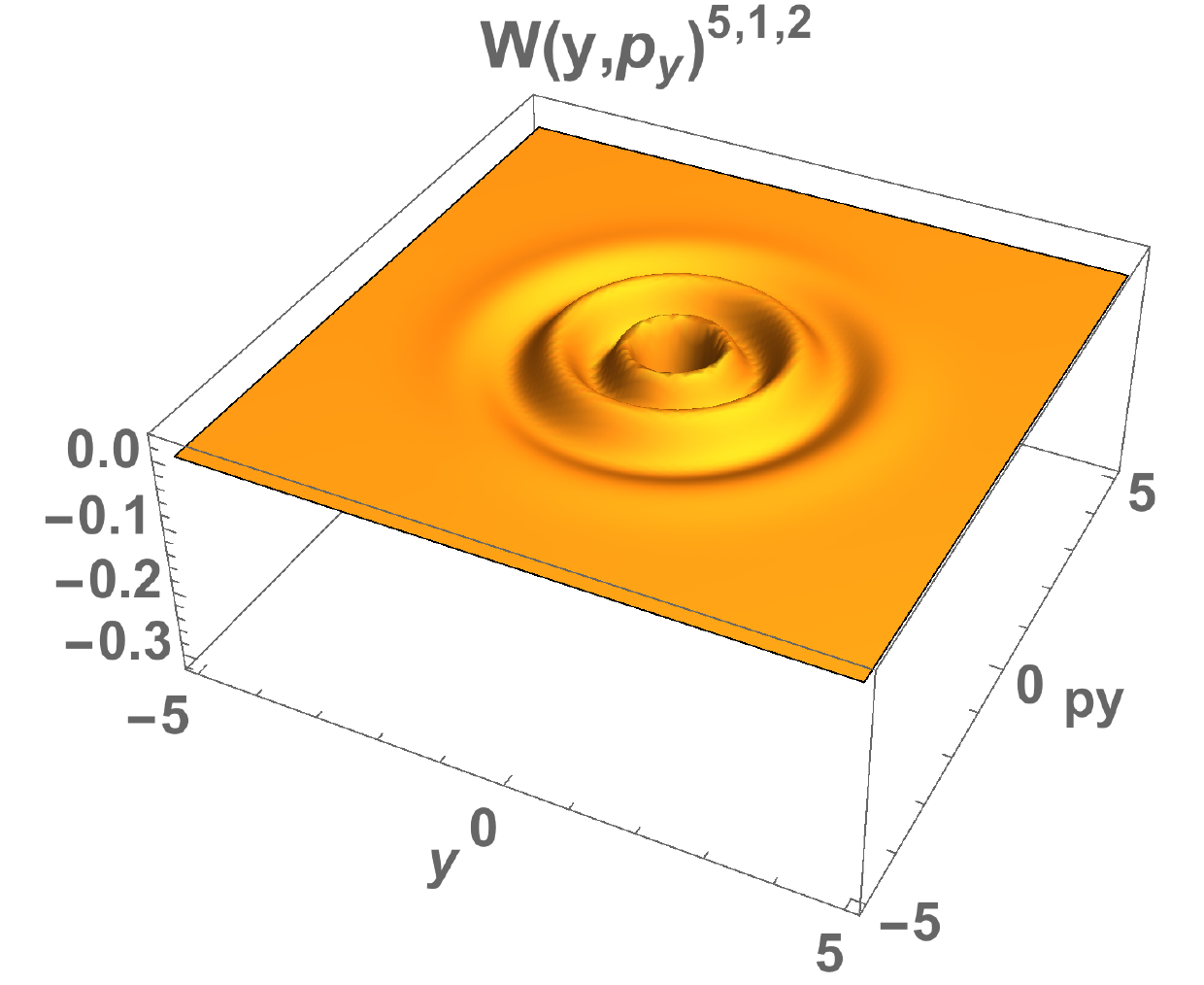}
\label{fig:WYPy}}
\qquad
\subfloat[$W(y,p_x)_{x=0,p_y=0}$]{
\includegraphics[scale=0.25]{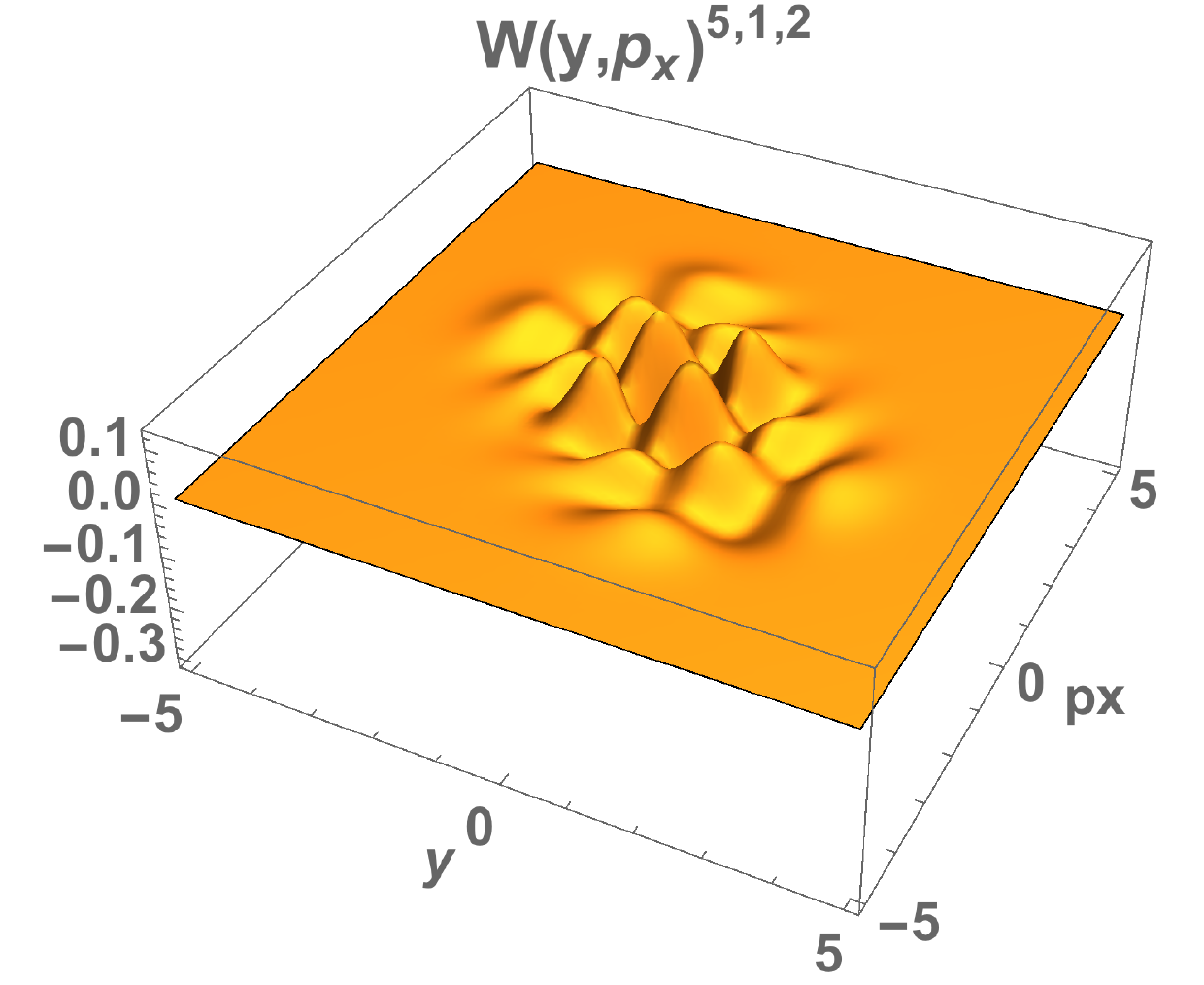}
\label{fig:WYPx}}
\caption{Wigner distribution of Ince-Gaussian vortex states for $p=5, m=1$ and $\epsilon=2$. Details are in the text.}
\label{fig:IGWigner}
\end{figure*}
It is thus observed that the order of the Ince-Gaussian vortex states, $p$ can be interpreted as the total number of photons $N$ in the two modes. Also, for a given order $p$, each value of the degree $m$ and ellipticity parameter $\epsilon$ gives rise to a different Ince-Gaussian vortex state comprising the same Laguerre-Gaussian vortex states. The general spatial quadrature distribution associated with Eq. (\ref{LGBasis}) can be derived from Eq. (\ref{LGquadrature}) and has the following form
\begin{eqnarray}
\label{IGVQuadrature}
\psi_v\left(r,\phi\right)&=&\sum_j^{\frac{N-1}{2}}A_j\sqrt{\frac{4j!}{\pi(N-j)!}}\left(\sqrt{2}r\right)^le^{i(N-2j)\phi}\nonumber \\
&\times & L_j^{N-2j}\left(2r^2\right)\exp\left(\frac{-r^2}{2}\right)
\end{eqnarray}
where the cylindrical polar coordinates are related to the spatial quadratures $x$ and $y$ by the regular transformation equations between cartesian and polar coordinates. In deriving Eq. (\ref{IGVQuadrature}) we have used $\psi\left(x,y\right)=\langle x,y\vert\psi\rangle_v$ and transformation property of Hermite and Laguerre polynomials.\\
We study the intensity distribution of the spatial quadratures of the Ince-Gaussian vortex states in Fig. \ref{fig:IGQuadratures} for different $p$, $m$ and $\epsilon$. In all the figures we see a vortex structure in the quadrature space. With increasing $p$, the central core increases in diameter. There are also two prominent crest regions and two troughs along the boundary of the central core. The height of the crests also increases with increasing $p$. Another noticeable feature is the disintegration of the boundary of the central core into separate lobes with increasing difference between $p$ and $m$. This means that the vortex nature depends of the closeness of values of $p$ and $m$. It is most prominent when $p=m$. Looking at figs (\ref{fig:IG330}), (\ref{fig:IG335}) and (\ref{fig:IG33Infinity}), we see that we get an uniform ring around the central core when ellipticity is either $0$ or $\infty$. This happens because when $\epsilon \rightarrow 0$, the Ince-Gaussian vortex state is transformed to a Laguerre-Gaussian vortex. On the other hand when $\epsilon \rightarrow \infty$, the Ince-Gaussian vortex is transformed to a Hermite-Gaussian vortex. Both Laguerre-Gaussian vortex and Hermite-Gaussian vortex are characterised by a uniform ring around the central core in the quadrature space. For all other intermediate values the ring is divided into two prominent crests and two prominent troughs.
\section{Entanglement and nonclassical properties}
\label{sec:Entanglement and nonclassical properties}
\begin{figure*}
\centering
\subfloat[$W(x,p_y)_{y=0,p_x=0}^{3,3,2}$]{
\includegraphics[scale=0.25]{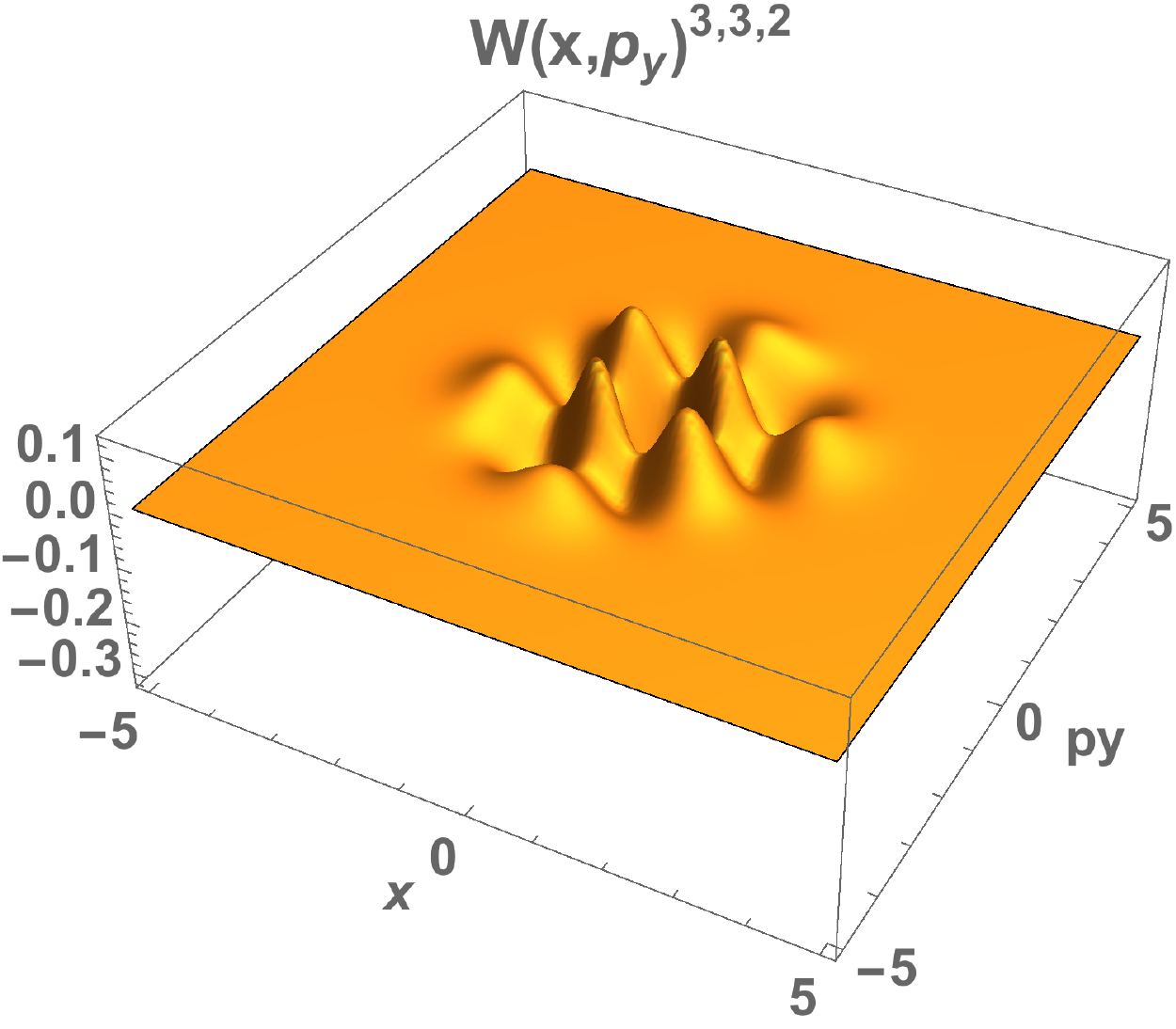}
\label{fig:W332}}
\qquad
\subfloat[$W(x,p_y)_{y=0,p_x=0}^{5,3,2}$]{
\includegraphics[scale=0.25]{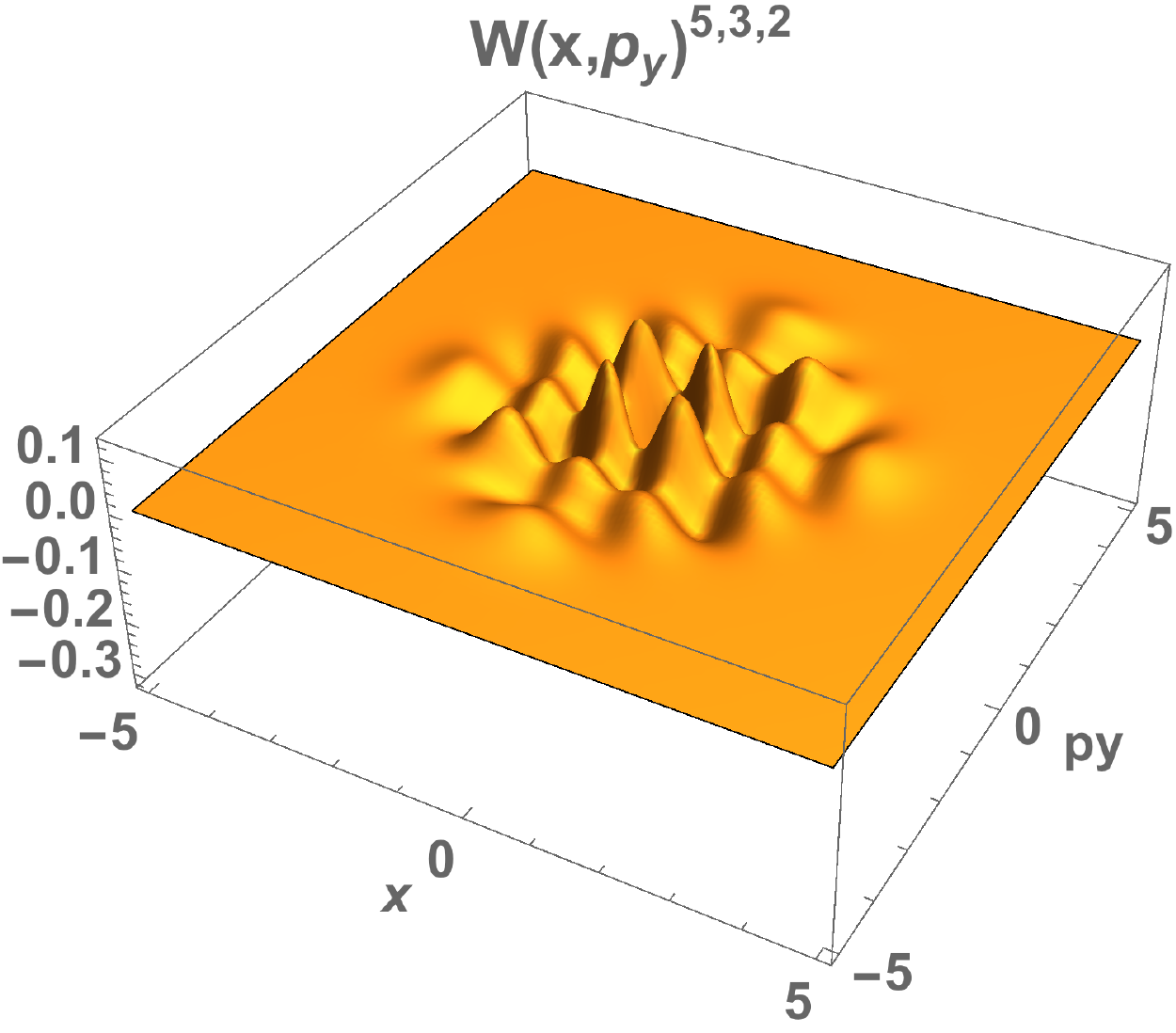}
\label{fig:W532}}
\qquad
\subfloat[$W(x,p_y)_{y=0,p_x=0}^{7,3,2}$]{
\includegraphics[scale=0.25]{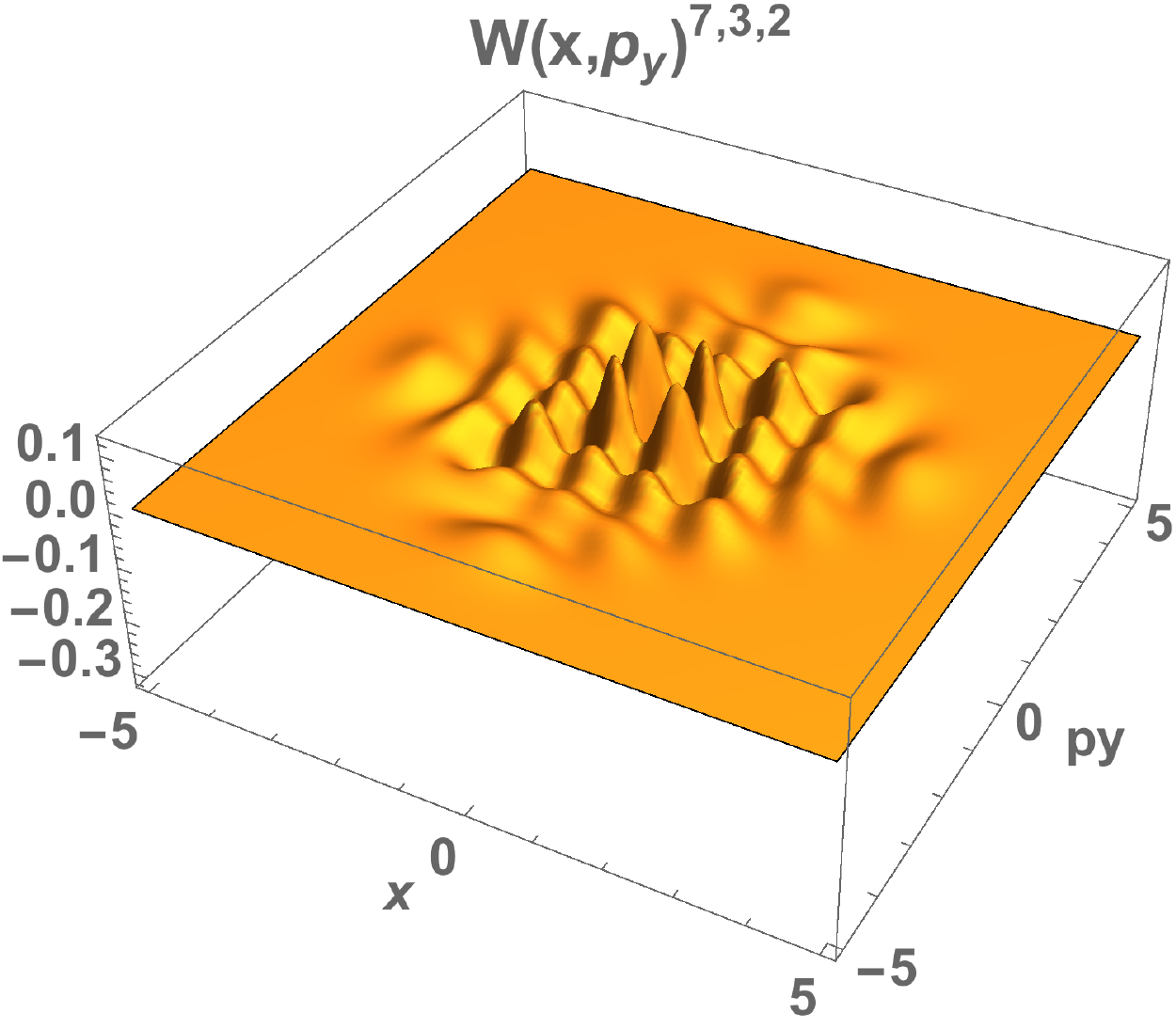}
\label{fig:W732}}
\qquad\\
\subfloat[$W(x,p_y)_{y=0,p_x=0}^{5,1,2}$]{
\includegraphics[scale=0.25]{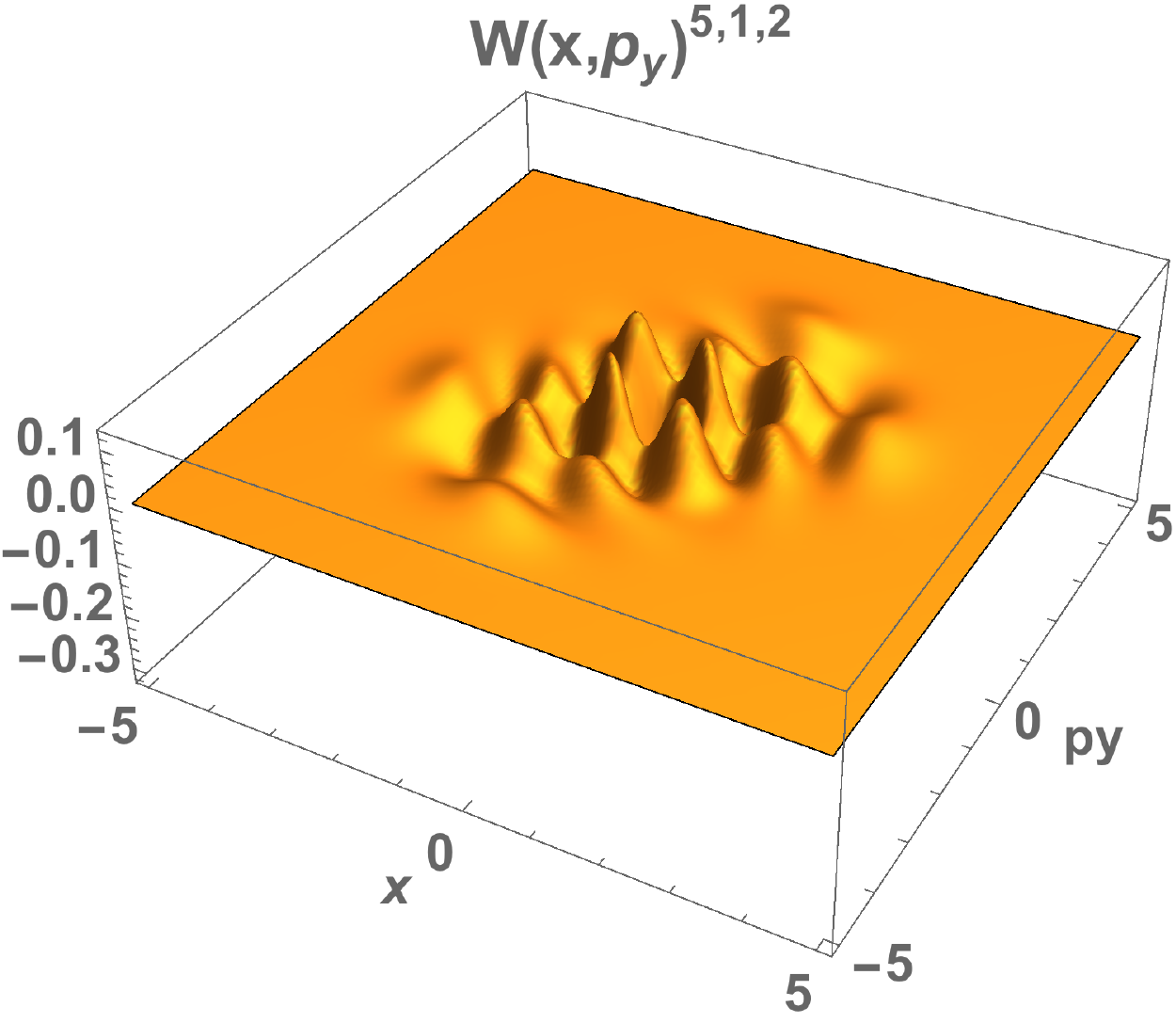}
\label{fig:W512}}
\qquad
\subfloat[$W(x,p_y)_{y=0,p_x=0}^{5,3,2}$]{
\includegraphics[scale=0.25]{W532.pdf}
\label{fig:W532a}}
\qquad
\subfloat[$W(x,p_y)_{y=0,p_x=0}^{5,5,2}$]{
\includegraphics[scale=0.25]{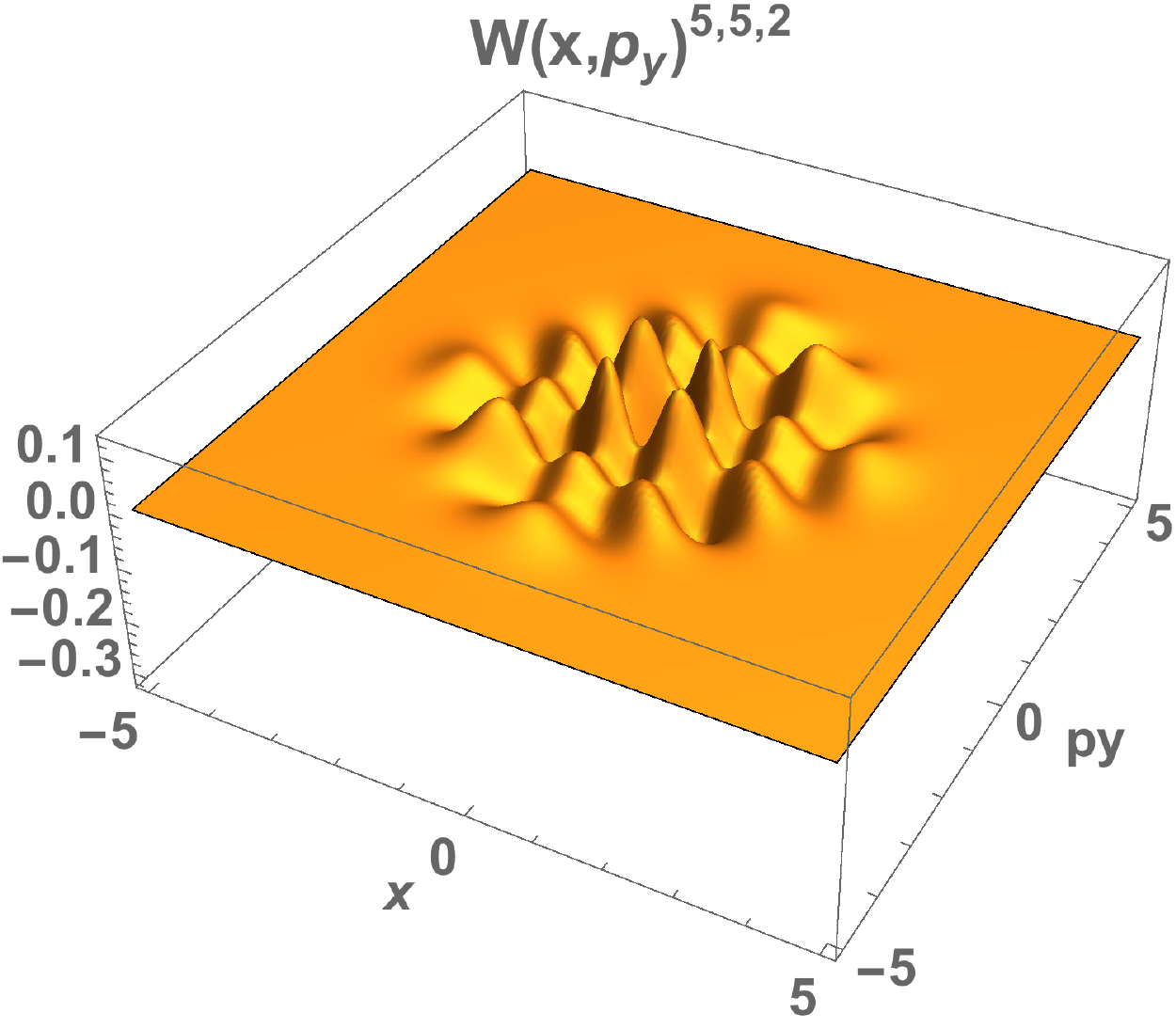}
\label{fig:W552}}
\caption{Wigner distribution of Ince-Gaussian vortex states for different $p$ and $m$. Details are in the text.}
\label{fig:IGWignerCombination}
\end{figure*}
In this section, we will look at the nonclassical properties of the quantum Ince-Gaussian vortex states with the help of the Wigner function. We will also study the variation of entanglement of quantum Ince-Gaussian vortex states with mode order and ellipticity parameter. We will use the von Neumann entropy for the same.\\
\subsection{Wigner function}
\label{WignerFunction}
In order to derive the Wigner distribution function \cite{wigner1932quantum} associated with the Ince-Gaussian vortex states, we will once again make use of the Laguerre-Gaussian basis. The Wigner function for Laguerre-Gaussian modes \cite{SimonAgarwal} with radial index $n$ and azimuthal index $l$ can be expressed in the following compact form
\begin{eqnarray}
\label{WignerLG}
W_{n,l}\left(\mathbf{r},\mathbf{p}\right)&=&\frac{(-1)^{2n+l}}{\pi}L_{n+l}\lbrace 4\left(Q_0+Q_1\right)\rbrace\nonumber\\
&\times & L_{n}\lbrace 4\left(Q_0-Q_1\right)\rbrace\exp\left(-4Q_0\right)
\end{eqnarray}
where $Q_0$ and $Q_1$ are quadratic functions which are described as follows
\begin{eqnarray}
\label{Quadratic}
Q_0 &=&\frac{1}{2}\left[\frac{x^2+y^2}{\sigma^2}+\frac{\sigma^2}{4}\left(p_x^2+p_y^2\right)\right],\\
Q_1 &=&\frac{xp_y-yp_x}{2}
\end{eqnarray}
Here, $x$ and $y$ are the spatial quadratures while $p_x$ and $p_y$ are the conjugate momentum quadratures. $L_i$ stands for Laguerre polynomial of radial index $i$. Following the above definition and using Eq. (\ref{LGBasis}) the Wigner function associated with the Ince-Gaussian vortex state of Eq. (\ref{FinalClosedForm}) can be written as
\begin{eqnarray}
\label{IGWigner}
W_{\psi_v}\left(\mathbf{r},\mathbf{p}\right)&=&\sum_{j}^{\frac{N-1}{2}}\vert A_j\vert^2\frac{(-1)^{N}}{\pi}L_{N-j}\lbrace 4\left(Q_0+Q_1\right)\rbrace\nonumber\\
&\times & L_{j}\lbrace 4\left(Q_0-Q_1\right)\rbrace\exp\left(-4Q_0\right)
\end{eqnarray}
We study $W_{\psi_v}\left(\mathbf{r},\mathbf{p}\right)$ in Fig. \ref{fig:IGWigner}. Since it is impossible to reproduce the entire Wigner function on paper and picturise it, we study all possible combinations of the four phase space quadratures taking two at a time while assuming the other two to be 0. We study the Wigner distribution as a function of two quadratures with $p$, $m$ and $\epsilon$ as parameters. It is observed from all the six combinations that there are negative regions. This implies the nonclassicality of the Ince-Gaussian vortex states. We see concentric rings with a central region of negative maximum in figs. (\ref{fig:WXY}), (\ref{fig:WXPx}), (\ref{fig:WYPy}) and (\ref{fig:WPxPy}). Interesting patterns appears in figs. (\ref{fig:WXPy}) and figs. (\ref{fig:WYPx}). These patterns are suggestive of quantum inteference between $\left(x,p_y\right)$ and $\left(y,p_x\right)$. Looking at these patterns more carefully in fig. (\ref{fig:IGWignerCombination}), we see that with increasing $p$, the spread of the pattern increases which implies increased interference between the two modes. If $p=m$, the interference pattern has a uniform spread. For all other combinations, the balance is lost and the pattern shrinks in one direction while spreading in the orthogonal direction.
\subsection{von Neumann Entropy}
\begin{figure}[!h]
\centering
\includegraphics[scale=0.4]{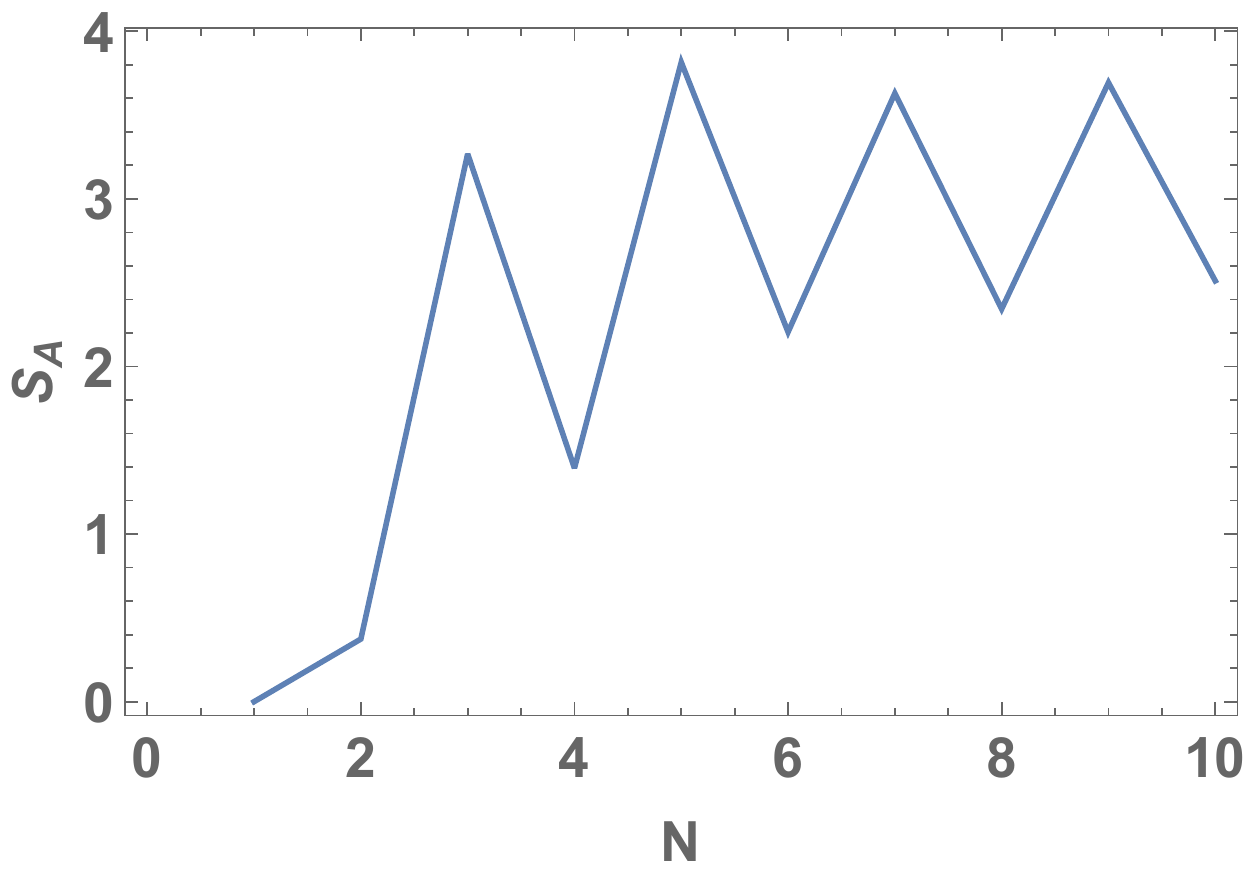}
\caption{Entropy variation with $N$.}
\label{fig:Entropy}
\end{figure}
One of the most trusted measures of pure state entanglement is the von Neumann entropy \cite{Nielsen}. For a two mode pure state, it is defined as
\begin{align}
\label{Entropy}
S_a &= -Tr\left(\rho_a\log\rho_a\right)\nonumber\\
&=-\sum_{n}\lambda_n\log\lambda_n
\end{align}
where $\rho_a$ is the reduced density matrix of the two mode state with density matrix $\rho_{ab}$ and $\lambda_n$ are the eigenvalues of the $\rho_a$. The density matrix associated with the Ince-Gaussian vortex state can be derived easily from Eq. (\ref{FinalClosedForm}). It has the form
\begin{align}
\label{DensityMatrix}
\rho_v &= \vert\psi\rangle_{v~v}\langle\psi\vert\nonumber\\
&= \sum_{j,j'}\frac{A_jA_j^*}{2^N}\sum_{k,k'}\sum_{l,l'}c_{lk}^{N}c_{l'k'}^{*Nj'}\nonumber\\
&\times \vert \psi_{jlk}\rangle\langle \psi_{j'l'k'}\vert
\end{align}
where $\vert\psi_{jlk}\rangle=\vert N-(j+l-k),j+l-k\rangle$. The reduced density matrix corresponding to the first mode is calculated from Eq. (\ref{DensityMatrix}) by taking a partial trace over the second mode. The resulting matrix is diagonalized numerically to calculate the eigenvalues. These eigenvalues are then used to calculate the von Neumann entropy of the Ince-Gaussian vortex states.\\
We study $S_a$ as a function of the total number of photons in Fig. \ref{fig:Entropy} with $m$ and $\epsilon$ as parameters. It is observed that the entanglement between two modes increases with increasing number of photons or increasing order in an oscillatory manner. This means that the entanglement for odd number of photons was always found to be more than the corresponding even number of photons. That is $S_a$ for $N=p=n$ is always greater than $S_a$ for $N=p=n-1$ or $N=p=n-1$ where $n$ is any odd number. The entanglement for even number of photons appears to reach a saturation level as the number of photons increases. The case for odd photon number is exactly similar but it saturates at a higher value compared to the former. Although it is difficult to exactly point out the reason behind this behavior, it appears that the expansion in Laguerre-Gauss vortex basis results in an unbalanced term in case of odd number of photons which results in a higher entanglement between the two modes.
\section{Conclusion}
\label{sec:Conclusion}
In this article we have presented a detailed theoretical method of generating quantum Ince-Gaussian vortex states. These states are comprised of photons exhibiting a complex vortex nature. The spatial quadrature distributions can be represented using Ince-Gaussian modes. These exotic modes are general solutions of the quantum harmonic oscillator in elliptical coordinates. Apart from the two quantum numbers that describe general quantum optical vortex states, an extra parameter is necessary to define the Ince-Gaussian vortex modes, the ellipticity parameter. It is the eccentricity of the elliptical coordinate system. For each value of this continuous parameter, there exists an orthogonal family of Ince-Gaussian modes.
We have observed that the order \emph{p} of the higher order vortex state is essentially the number of photons with both modes combined. In our study of the quadrature distribution, it was seen that the condition $p=m$ resulted in greater stability of the vortex state. This means that the central core of zero intensity was bounded by a region of uniform intensity. With increasing difference between \emph{p} and \emph{m}, this region of uniform intensity starts to split into two separate lobes. Similarly in case of the Wigner distribution, $p=m$ resulted in an interference pattern with uniform spread in all directions. For all other cases, the interference pattern exhibits a squeezed distribution. It should be noted that all six combinations of the Wigner distribution exhibited negative regions confirming the nonclassical nature of the state under study.\\
An interesting observation was made in our analysis of the entanglement of the higher order vortex state. We observed that in general entanglement for the vortex state with odd \emph{p} and \emph{m} was higher compared to its immediate even neighbors (even \emph{p} and \emph{m}). Overall though, it was observed that the entanglement appeared to saturate with increasing number of photons.
\acknowledgments{This work is partially supported by DST through SERB grant no.: SR/S2/LOP - 001/2014.

\end{document}